\documentclass[a4paper,12pt]{article}
\usepackage[latin1]{inputenc}
\usepackage[T1]{fontenc}
\usepackage{amsmath, amsthm, amssymb}
\usepackage{array}

\usepackage[final]{graphicx}   
\graphicspath{{tikz/}}

\newtheorem{definition}{Definition}[section]
\newtheorem{theorem}[definition]{Theorem}
\newtheorem{lemma}[definition]{Lemma}

\newtheorem{example}[definition]{Example}

\begin{document}

\pagenumbering{roman}
\pagestyle{empty}
\title{Constructions of MDS convolutional codes using superregular matrices}
\author{Julia Lieb and Raquel Pinto}
\maketitle




\setcounter{page}{1}


\pagenumbering{arabic}
\pagestyle{plain}

\begin{abstract}
Maximum distance separable convolutional codes are the codes that present best performance in error correction among all convolutional codes with certain rate and degree. In this paper, we show that taking the constant matrix coefficients of a polynomial matrix as submatrices of a superregular matrix, we obtain a column reduced generator matrix of an MDS convolutional code with a certain rate and a certain degree. We then present two novel constructions that fulfill these conditions by considering two types of superregular matrices.
\end{abstract}

\section{Introduction}

The (free) distance of a code measures its capability of detecting and correcting errors introduced during information transmission through a noisy channel. Maximum distance separable (MDS) codes are the ones that have maximum distance among all codes with the same parameters. MDS block codes of rate $k/n$ are the block codes with distance equal  to the Singleton bound $n-k+1$. The class of MDS block codes is very well understood and there exist relevant MDS block codes like the Reed-Solomon codes \cite{MacWilliams1988bk}.

The theory of convolutional codes is more involved and there are not many known constructions of MDS convolutional codes. Maximum distance separable (MDS) codes have maximum free distance in the class of convolutional codes of a certain rate $k/n$ and a certain degree $\delta$, i.e., are the ones with free distance equal to the Singleton bound $(n-k)\left(\left\lfloor \frac{\delta}{k} +1\right\rfloor  \right) + \delta +1$ \cite{MDS}. The first construction of MDS convolutional codes was obtained by Justesen in \cite{ju75} for codes of rate $1/n$ and restricted degrees. In \cite{sm98p1} Smarandache and Rosenthal presented constructions of convolutional codes of rate $1/n$ and arbitrary degree using linear systems representations. However these constructions require a larger field size than the constructions obtained in \cite{ju75}. Gluesing-Luerssen and Langfeld introduced in \cite{gll06} a new construction of convolutional codes of rate $1/n$ that requires the same field sizes as the ones obtained in \cite{ju75} but also with a restriction on the degree of the code. Finally, Smarandache, Gluesing-Luerssen and Rosenthal \cite{sm01a} constructed MDS convolutional codes for arbitrary parameters.

We will define a new construction of convolutional codes of any degree and sufficiently low rate using superregular matrices with a specific property. We then provide explicitly constructions of these codes using Cauchy circulant matrices \cite{Roth1989} and superregular matrices as defined in \cite{dr16}. A similar procedure was done for constructing two-dimensional MDS convolutional codes \cite{2D,k1}.

The paper is organized as follows: In the next section we start by introducing some preliminaries on superregular matrices. We give the definition of these matrices and two different types of superregular matrices. then we give some definitions and results on convolutional codes. In Section \ref{conditions}, we present a procedure to construct MDS convolutional codes using superegular matrices. We show that generator matrices whose coefficients of its entries fulfill certain conditions are generator matrices of an MDS convolutional code. In Section \ref{sec:constructions}, we give two different constructions of MDS convolutional codes of an arbitrary degree and rate smaller than some upper bound. Finally, in Section \ref{sec:comparison}, we compare the necessary field size and the restrictions on the parameters of our obtained constructions with those of already known constructions.

\section{Preliminaries}

\subsection{Superregular matrices}

\begin{definition}[\cite{Roth1989}] \label{def2}
A matrix $A \in \mathbb F^{r \times \ell}$ is said to be \textbf{superregular} if every minor of $A$ is different from zero.
\end{definition}

The following lemma is easy to see and we will use it several times to derive our conditions for MDS convolutional codes.

\begin{lemma}\label{le2}
(i) Let $A \in \mathbb F^{r \times \ell}$ be superregular. Then, each vector which is a linear combination of $s$ columns of $A$ has at most $s-1$ zeros.\\
(ii) Let $A \in \mathbb F^{r \times \ell}$ with $r\geq\ell$ be such that all its fullsize minors are nonzero. Then, each vector which is a linear combination of $\ell$ columns of $A$ has at most $\ell-1$ zeros.
\end{lemma}

There are many examples of superregular matrices. We will present two types of superregular matrices that we will use later for the constructions that we  introduce in this paper. The first one will be the Cauchy circulant matrices.

\begin{theorem}\cite{Roth1989} \label{cauchy}
Let $\mathbb{F}$ be a finite field with $|\mathbb F|=q$ where $q$ is an odd number. Furthermore, let $\alpha$ be an element of order $\frac{(q-1)}{2}$ and let $b$ be a nonsquare element in $\mathbb{F}$.
Then the $(\frac{q-1}{2}) \times (\frac{q-1}{2})$ matrix $C = \big[ \, c_{ij} \, \big]$ where
\[
  c_{ij}
  =
  \frac{1}{1-b\alpha^{j-i}},
  \quad \text{for} \quad
  0 \leq i,j \le \frac{q-3}{2}
\]
is superregular.
\end{theorem}

The matrix considered in the above theorem is a Cauchy circulant matrix. Another type of superregular matrix is given in the next theorem.

\begin{theorem}\cite{dr16}\label{diagonal} Let $\alpha$ be a primitive element of a finite field $\mathbb{F}=\mathbb{F}_{p^N}$ and $B=[\nu_{i\,\ell}]$ be a matrix over $\mathbb{F}$ with the following properties
\begin{enumerate}
\item $\nu_{i\,\ell}=\alpha^{\beta_{i\,\ell}}$ for a positive integer $\beta_{i\,\ell}$;
\item if $\ell<\ell'$, then $2\beta_{i\,\ell}\leq\beta_{i\,\ell'}$;
\item if $i<i'$, then $2\beta_{i\,\ell}\leq\beta_{i'\,\ell}$.
\end{enumerate}
Suppose $N$ is greater than any exponent of $\alpha$ appearing as a nontrivial term of any minor of $B$. Then $B$ is superregular.
\end{theorem}

\subsection{Convolutional codes}

Let $\mathbb F$ be a finite field and $\mathbb F[z]$ the ring of the polynomials with coefficients in $\mathbb F$. A \textbf{convolutional code of rate $k/n$} is an $\mathbb F[z]$-submodule of $\mathbb F[z]^n$ of rank $k$. An \textbf{generator matrix} of a convolutional code $\mathcal{C}$ of rate $k/n$ is any $n \times k$ matrix whose columns constitute a basis of $\mathcal C$, i.e., it is a full column rank matrix $G(z)$ such that
\begin{eqnarray*}
  \mathcal{C} &=& Im_{\mathbb F[z]}G(z) \\
  &=& \{G(z)u(z) \, : \, u(z) \in \mathbb F[z]^k\}.
\end{eqnarray*}
 If $G(z) \in \mathbb F[z]^{n \times k}$ is a generator matrix of a convolutional code $\mathcal{C}$, then all generator matrices of $\mathcal{C}$ are of the form $G(z)U(z)$ for some unimodular matrix $U(z) \in \mathbb F[z]^{k \times k}$. Two generator matrices of the same code are said to be \textbf{equivalent generator matrices}.

 Since two equivalent generator matrices differ by right multiplication of a unimodular matrix, they have the same full size minors, up to multiplication by a nonzero constant. The \textbf{complexity} or \textbf{degree} of a convolutional code is defined as the maximum degree of the full size minors of a generator matrix of the code.

 Define the j-th column degree $\nu_j$ of a polynomial matrix $G(z) \in \mathbb F[z]^{n \times k}$ to be the maximum degree of the entries of the j-th column of $G(z)$. Obviously, the sum of the column degrees of $G(z)$ is greater or equal than the maximum degree of its full size minors. If the sum of the column degrees of $G(z)$ equals the maximum degree of its full size minors, $G(z)$ is said to be \textbf{column reduced}. A convolutional code always admits column reduced generator matrices and two column reduced generator matrices have the same column degrees up to a column permutation \cite{Forney73,Kailath80}. Therefore, column reduced generator matrices are the ones that have minimal sum of the column degrees and such the sum of its column degrees is equal to the degree of the code.

\begin{definition}
For $G(z)\in\mathbb F[z]^{n\times k}$, let $[g_{ij}]_{hc}$ denote the coeffcient of $z^{\nu_j}$ in $g_{ij}(z)$. Then, the \textbf{highest column degree coeffcient matrix} $[G]_{hc}\in\mathbb F^{n\times k}$ is defined as the matrix
consisting of the entries $[g_{ij}]_{hc}$.
\end{definition}
It holds that $G(z)$ is column reduced if and only if $[G]_{hc}\in\mathbb F^{n\times k}$ has full rank.

The \textbf{weight} of a vector $c \in \mathbb F^n$, $wt(c)$, is the number of its nonzero entries and the weight of a polynomial vector $v(z)=\displaystyle\sum_{i\in\mathbb N_0} v_iz^i \in \mathbb F[z]^n$ is given by
\[
wt(v(z))=\displaystyle \sum_{i \in \mathbb N_0} wt(v_i).
\]
The \textbf{free distance} of a convolutional code $\mathcal{C}$ is the minimum weight of the nonzero codewords of the code, i.e.,
\[
d_{free}(\mathcal{C})=\{wt(v(z)) \,: \, v(z) \in \mathcal{C} \backslash \{0\} \}.
\]

In \cite{MDS} Smarandache and Rosenthal obtained an upper bound for the free distance of a convolutional code $\mathcal{C}$ of rate $k/n$ and degree $\delta$ given by
\[
d_{free}(\mathcal{C}) \leq (n-k) \left(\left\lfloor \frac{\delta}{k} \right\rfloor +1  \right) + \delta +1.
\]
This bound is called the \textbf{generalized Singleton bound}.
A convolutional code of rate $k/n$ and degree $\delta$ with free distance equal to the generalized Singleton bound is called \textbf{Maximum Distance Separable (MDS)} convolutional code. If $\mathcal{C}$ is such a code and $G(z) \in \mathbb F[z]^{k \times n}$ is a column reduced generator matrix of $\mathcal{C}$, its columns degrees are equal to $\left\lfloor \frac{\delta}{k} \right\rfloor +1 $ with multiplicity $t:=\delta - k \left\lfloor \frac{\delta}{k} \right\rfloor$ and $\left\lfloor \frac{\delta}{k} \right\rfloor $ with multiplicity $k-t$; see \cite{MDS}.

\section{Conditions to obtain MDS convolutional codes}\label{conditions}

Let $\mathcal{C}$ be a convolutional code of rate $k/n$ and degree $\delta$. In this section, we will derive conditions on a column reduced generator matrix $G(z)$ of $\mathcal{C}$ that ensure that the code is an MDS convolutional code.




To this end, we assume that $G(z)$ has non-increasing column degrees with values $\left\lfloor \frac{\delta}{k} \right\rfloor +1 $ and $\left\lfloor \frac{\delta}{k} \right\rfloor $. We write $G(z)=\sum_{i=0}^{\mu}G_iz^{i}$ with $G_{\mu}\neq 0$, i.e. $\mu=\deg G$, and $\nu:=\left\lfloor \frac{\delta}{k} \right\rfloor + 1
$, i.e. $\nu=\mu$ if $k\nmid \delta$ and $\nu=\mu+1$ if $k\mid\delta$.

Furthermore, we write $G(z)=[g_1(z)\hdots g_k(z)]$ with $$g_r(z)=\begin{cases}  \displaystyle
    \sum_{0 \leq i \leq \nu}
   g_{i,r} z^{i},
      &   \text{for $r = 1, 2, \ldots, t$}, \\
      \displaystyle
    \sum_{0 \leq i \leq \nu-1}
   g_{i,r} z^{i},
      &  \text{for $r = t+1, t+2, \ldots, k$},\\
      \end{cases}$$

      i.e. $G_i=[g_{i,1}\cdots g_{i,k}]$ for $i=1,\hdots,\nu-1$ and $G_{\nu}=[g_{\nu,1} \cdots g_{\nu,t}\ 0\cdots 0]$, where $t=\delta - k \left\lfloor \frac{\delta}{k} \right\rfloor$.
Set
\begin{align}\label{g} \mathcal{G}
    = \left[
          \begin{array}{cccccc}
            g_{0,1} \cdots  g_{\nu,1} & \cdots & g_{0,t}  \cdots g_{\nu,t} & g_{0,t+1}  \cdots g_{\nu-1,t+1} &  \cdots & g_{0,k}  \cdots  g_{\nu-1,k}
          \end{array}
        \right]
        \in \mathbb{F}^{n \times (k\nu + t)}.
\end{align}
Write $u(z)=\sum_{i\in\mathbb N_0}u_iz^{i}$.\\
If $l:=\deg{u}\leq \mu$, it holds
\begin{align}\label{1}
&v(z)=G(z)u(z)\nonumber\\
&=G_0u_0+\cdots [G_l\cdots G_0]\begin{pmatrix}u_0\\ \vdots\\u_l\end{pmatrix}z^l+\cdots+[G_{\mu}\cdots G_{\mu-l}]\begin{pmatrix}u_0\\ \vdots\\u_l\end{pmatrix}z^{\mu}+\cdots+G_{\mu}u_lz^{\mu+l}
\end{align}
For $l\geq\mu$, one obtains
\begin{align}\label{2}
&v(z)=G(z)u(z)\nonumber\\
&=G_0u_0+\cdots +[G_{\mu}\cdots G_0]\begin{pmatrix}u_0\\ \vdots\\u_{\mu}\end{pmatrix}z^{\mu}+\cdots+[G_{\mu}\cdots G_0]\begin{pmatrix}u_{l-\mu}\\ \vdots\\u_l\end{pmatrix}z^{l}+\cdots+G_{\mu}u_lz^{\mu+l}
\end{align}

As $wt(G(z)u(z))=wt(G(z)u(z)z^r)$ for $r\in\mathbb N$, throughout this paper, we assume, without loss of generality that $u_0\neq 0$.

\begin{theorem}
If the matrix $\mathcal{G}$ defined in \eqref{g} is superregular, $G(z)$ is the generator matrix of an $(n,k,\delta)$ convolutional code.
\end{theorem}

\begin{proof}
Since the highest column degree coefficient matrix of $G(z)$ is equal to
\[
  \left[
    \begin{array}{ccccccc}
      g_{\nu,1} & g_{\nu,2} & \cdots & g_{\nu,t} & g_{\nu-1, t+1} & \cdots & g_{\nu-1,k}
    \end{array}
  \right],
\]
it is a submatrix of the superregular matrix $\mathcal{G}$ and hence full rank. Consequently, $G(z)$ is column reduced. Therefore, the degree of the code generated by $G(z)$ is equal to the sum of the column degrees of $G(z)$, which is $\nu t + (\nu-1)(k-t)=\delta$.
\end{proof}

The generated code is an MDS convolutional code if and only if for each $u(z)\in\mathbb F[z]^k\setminus\{\mathbf{0}\}$ and $v(z)=G(z)u(z)$, it holds
\begin{equation}\label{wt}
wt(v(z))\geq (n-k)
  \left( \left\lfloor \frac{\delta}{k} \right\rfloor + 1 \right)
  +  \delta  +  1=n \nu - (k-t) + 1.
   \end{equation}

Next, we will show that under certain conditions equation \eqref{wt} is fulfilled by considering different cases for the value of $\delta$. In any case, one of the conditions will always be the superregularity of $\mathcal{G}$. However, this condition is not necessary to obtain an MDS convolutional code as the following example shows.

\begin{example}\ \\
Let $\mathcal{C}$ be the convolutional code of rate $2/3$ and degree $1$ with generator matrix $G(z)=\left[\begin{array}{cc} 1 & 1\\ 1& 2\\0 & 1\end{array}\right]+\left[\begin{array}{cc} 1 & 0\\ 1& 0\\ 2 & 0\end{array}\right]z$.
The free distance of this code is $d_{free}(\mathcal{C})=3$ and hence it is an MDS convolutional code but $\mathcal{G}=\left[\begin{array}{ccc} 1 & 1 &1\\ 1& 2 & 1\\0 & 1 & 2\end{array}\right]$ is not superregular.
\end{example}

\subsection{Conditions for the case $\delta< k$}\ \\
In this case, we have to prove that $wt(v(z))\geq n-k+\delta +1$.

\begin{theorem} \label{th2}
Assume that $\delta< k$ and let $\mathcal{G}$ be superregular.
If $n\geq \delta+k-1$,
then $G(z)$ is the generator matrix of an $(n,k,\delta)$ MDS convolutional code.
\end{theorem}

\begin{proof}
As $\delta<k$, it holds $\nu=\mu=1$ and $t=\delta$.\\
\textbf{Case 1:} $l=0$\\
One has $v(z)=G_0u_0+G_1u_0z$, where $G_0$ and the $\delta$ nonzero columns of $G_1$ form superregular matrices. If the first $\delta$ components of $u_0$ are zero, i.e. $G_1u_0=0$, it holds $wt(v(z))\geq n-(k-\delta)+1$ since $G_0u_0$ is a nonzero linear combination of $k-\delta$ columns of $G_0$. If one of the first $\delta$ components of $u_0$ is nonzero, one obtains $wt(v(z))=wt(G_0u_0)+wt(G_1u_0)\geq n-k+1+n-\delta+1\geq n-k+\delta+1$ as $n\geq\delta+k-1\geq 2\delta-1$.\\
\textbf{Case 2:} $l\geq 1$\\
Here, one has $v(z)=G_0u_0+[G_1\ G_0]\begin{pmatrix}u_0\\ u_1\end{pmatrix}z+\cdots+[G_1\ G_0]\begin{pmatrix}u_{l-1}\\ u_l\end{pmatrix}z^{l}+G_1u_lz^{l+1}$. If the first $\delta$ components of $u_l$ are zero, it holds\\ $wt(v(z))\geq wt(G_0u_0)+wt\left([G_1\ G_0]\begin{pmatrix}u_{l-1}\\ u_l\end{pmatrix}\right)\geq n-k+1+n-(k+\delta-\delta)+1\geq n-k+\delta+1$ since $[G_1\ G_0]\begin{pmatrix}u_{l-1}\\ u_l\end{pmatrix}$ is a nonzero linear combination of $\delta$ columns of $G_1$ and $k-\delta$ columns of $G_0$ and $n\geq \delta+k-1$. If one of the first $\delta$ components of $u_l$ is nonzero, one obtains $wt(v(z))\geq wt(G_0u_0)+wt(G_1u_l)\geq n-k+1+n-\delta+1\geq n-k+\delta+1$ as $n\geq\delta+k-1\geq 2\delta-1$.
%
\end{proof}

\subsection{Conditions for the case $\delta\geq k$}\ \\
\noindent For this subsection, we need the additional definitions\\
$\mathcal{G}_1=\begin{pmatrix} G_0\\ \vdots\\ G_\nu \end{pmatrix}\in\mathbb F^{(\nu+1)n\times k}$, $\mathcal{G}_2=\begin{pmatrix} G_0\\ \vdots\\ G_{\nu-1}\end{pmatrix}\in\mathbb F^{\nu n\times k}$,
$\bar{\mathcal{G}}=\begin{pmatrix} G_0\\ \vdots\\ G_\mu \end{pmatrix}\in\mathbb F^{(\mu+1)n\times k}$.

\noindent
It holds $\bar{\mathcal{G}}=\mathcal{G}_1$ for $k\nmid\delta$ and $\bar{\mathcal{G}}=\mathcal{G}_2$ for $k\mid\delta$ and
\begin{align}\label{I}
G(z)=[I_n\ I_nz\cdots I_nz^{\mu}]\bar{\mathcal{G}}.
\end{align}

\begin{theorem} \label{th2}
Assume that $\delta\geq k$ and let $\mathcal{G}$ defined in \eqref{g} be superregular. Moreover, assume that all fullsize minors of $\bar{\mathcal{G}}$ are nonzero.
If $n\geq 2\delta+k-\nu$,
then $G(z)$ is the generator matrix of an $(n,k,\delta)$ MDS convolutional code.
\end{theorem}

\begin{proof}

We distinguish several cases.\\
\textbf{Case 1}: $l=0$\\
\textbf{Case 1.1}: $k\mid\delta$\\
In this case, the generalized Singleton bound is equal to $n\nu-k+1$.
If we define $v=\mathcal{G}_2u$, we obtain that $v$ is a nonzero linear combination of the columns of a matrix with nonzero fullsize minors and hence $wt(v)\geq n\nu-k+1$.

\noindent\textbf{Case 1.2}: $k\nmid\delta$

Let us write
\begin{math}
  u_0
  =
  \left[
    \begin{array}{c}
    u_0^{(1)} \\
      u_0^{(2)}
    \end{array}
  \right],
\end{math}
with $u_0^{(1)} \in \mathbb{F}^{t}$ and $u_0^{(2)} \in \mathbb{F}^{k-t}$, and set $v = \mathcal{G}_1 u$.
Then
\begin{math}
  v
  =
  \left[
    \begin{array}{c}
    v^{(1)} \\
      v^{(2)}
    \end{array}
  \right],
\end{math}
with $v^{(1)}=\mathcal{G}_2u\in\mathbb F^{\nu n}$ and $v^{(2)}=G_{\nu}u^{(1)}\in\mathbb F^{n}$.

Hence, $v^{(1)}$ is a nontrivial linear combination of columns of an $n \nu \times k$ matrix with nonzero fullsize minors and $v^{(2)}$ is a linear combination of columns of an $n \times t$ matrix with nonzero fullsize minors.
We distinguish two further subcases.

\textbf{Case 1.2.1:} $u^{(1)} =0$.
In this case, one has
\begin{math}
  v
  =
  \left[
    \begin{array}{c}
      v^{(1)} \\
      0
    \end{array}
  \right]
\end{math}
where $v_{1}$ is a nontrivial linear combination of the columns of an $n \nu \times (k-t)$ matrix with nonzero fullsize minors and $k-t < n \nu$.
Applying Lemma \ref{le2}, we obtain $wt(v) \geq n \nu - (k-t) + 1$.

\textbf{Case 1.2.2:} $u^{(1)} \neq 0$.
In this case, $v^{(1)}$ and $v^{(2)}$ are nontrivial linear combinations of the columns of an $n \nu \times k$ and an $n  \times t$ matrix with nonzero fullsize minors, respectively.
Moreover, since $n \nu > k$ and $n > t$, it follows from Lemma \ref{le2} that $wt(v^{(1)}) \geq n \nu - k + 1$ and $wt(v^{(2)}) \geq n - t + 1$ and thus we get
\begin{align*}
 wt(v)
  & =
 wt(v^{(1)}) +wt(v^{(2)}) \\
  & \geq
  n \nu+n - k - t + 2
  =
  n \nu - (k-t) + 1 +n-2t+1\geq  n \nu - (k-t) + 1
\end{align*}
where the last inequality follows from the fact that $n\geq 2\delta+k-\nu=\delta+k-1+\delta-\lfloor\frac{\delta}{k}\rfloor\geq \delta+k-1\geq 2t-1$.

Using equation (\ref{I}), $wt(G(z) u(z)) = wt(v)$ and the result follows.

\noindent\textbf{Case 2:} $1\leq l< \mu$\\
Note that for this case, one has
\begin{align}\label{eqn}
n\geq 2\delta+k-\nu\geq k+\delta-1=\left( \frac{\delta}{k} -\frac{1}{k}+ 1\right)k\geq \mu k\geq (l+1)k.
\end{align}
\textbf{Case 2.1:} $k\mid\delta$\\
Using equations \eqref{1} and \eqref{eqn}, the superregularity of $\mathcal{G}$ and that $u_0$ and $u_l$ are nonzero, we obtain
\begin{align*}
&wt(v(z))\geq 2 \sum_{i=1}^l(n-ik+1) +(\frac{\delta}{k}-l+1)(n-(l+1)k+1)=\\
&=2nl+2l-k(l+1)l+(\frac{\delta}{k}+1)(n-k)+(\frac{\delta}{k}+1)(-lk+1)-ln+l(l+1)k-l\\
&=(\frac{\delta}{k}+1)(n-k)+\delta+1+nl+l+(\frac{\delta}{k}+1)(-lk+1)-\delta-1.
\end{align*}
Consequently, in order to get $wt(v(z))\geq(\frac{\delta}{k}+1)(n-k)+\delta+1$, one needs
$$n\geq \frac{1}{l}\left(\delta+1-l+l\delta+lk-\frac{\delta}{k}-1\right)=\delta+k-1+\frac{1}{l}\left(\delta-\frac{\delta}{k}\right).$$
The result follows from $\delta+k-1+\frac{1}{l}\left(\delta-\frac{\delta}{k}\right)\leq 2\delta+k-\nu$.\\
\textbf{Case 2.2:} $k\nmid\delta$\\
Additionally to the previous subcase, here we have to regard that $G_{\mu}u_l$ might be zero and that $[G_{\mu}\cdots G_i]$ for $i={\mu-l},\hdots \mu-1$ has $k-t=k\mu-\delta$ zero columns. Therefore, we get
\begin{align*}
&wt(v(z))\\
&\geq 2 \sum_{i=1}^l(n-ik+1)-(n-k+1) +(\mu-l+1)(n-(l+1)k+1)+(k\mu-\delta) l\\
&=\mu(n-k)+\delta+1+nl+l-lk+\mu(-lk+1)+(k\mu-\delta) l-\delta-1\\
&=\mu(n-k)+\delta+1+nl+l-lk+\mu-\delta l-\delta-1
\end{align*}
Hence, one needs $n\geq k+\delta-1+\frac{1}{l}(\delta+1-\mu)$. This holds as $k+\delta-1+\frac{1}{l}(\delta+1-\mu)\leq k+2\delta-\mu$ and $\mu=\nu$ for $k\nmid\delta$.\\
\textbf{Case 3:} $l\geq\mu$\\
For this case, we consider equation \eqref{2}. As it could happen that $2\delta+k-\nu$ is smaller than $(\mu+1)k$, the weight of $v_i$ might be zero for $i=\mu,\hdots, l$. However, it holds $\mu k\leq \left( \frac{\delta}{k} -\frac{1}{k}+ 1\right)k=\delta+k-1\leq 2\delta+k-\nu$.\\
\textbf{Case 3.1:} $k\mid\delta$\\
Using equation \eqref{2} and the superregularity of $\mathcal{G}$, we obtain
\begin{align*}
&wt(v(z))\geq 2 \sum_{i=1}^{\mu}(n-ik+1)=2n\mu+2\mu-(\mu+1)\mu k\\
&=(n-k)(\mu+1)+n(\mu-1)+2\mu-(\mu+1)(\mu-1)k+\delta+1-\delta-1.
\end{align*}
Hence, for $\mu\geq 2$, one needs $n\geq k(\mu+1)-2+\frac{\delta-1}{\mu-1}=k+\delta-2+\frac{\delta-1}{\mu-1}$.\\
This is true for $\mu\geq 3$ since $k+\delta-2+\frac{\delta-1}{\mu-1}\leq k+\frac{3}{2}\delta-\frac{5}{2}\leq k+2\delta-\nu$ for $k\geq 2$ and $k+\delta-2+\frac{\delta-1}{\mu-1}=k+\delta-1$ for $k=1$. It is also true for $\mu=2$ since $k+\delta-2+\delta-1=k+2\delta-3\leq k+2\delta-\nu$.\\
It remains to consider the case $\mu=1$.\\
If we consider above estimation for the weight for $\mu=1$, we get the condition $\delta\leq 1$. Hence, for the following consideration, we could assume $k=\delta\geq 2$. For these parameters it holds $2\delta+k-\nu\geq k+\delta$ and thus, we could exploit the superregularity of $[G_1\ G_0]$. Doing this, we get
\begin{align*}
wt(v(z))&\geq 2(n-k+1)+(n-2k+1)=2(n-k)+\delta+1-\delta+2+n-2k\\
&\geq 2(n-k)+\delta+1
\end{align*}
because $n\geq 2\delta+k-\nu=\delta+2k-2$.\\
\textbf{Case 3.2:} $k\nmid\delta$\\
If one of the first $t$ components of $u_l$ is nonzero, we get
\begin{align*}
wt(v(z))\geq 2 \sum_{i=1}^{\mu}(n-ik+1)
=(n-k)\mu+n\mu+2\mu-\mu^2k+\delta+1-\delta-1.
\end{align*}
Consequently, one needs $n\geq k\mu-2+\frac{\delta+1}{\mu}$, which is true as\\ $k\mu-2+\frac{\delta+1}{\mu}\leq k+\delta-3+\frac{\delta+1}{2}\leq k+\frac{3}{2}\delta-\frac{5}{2}\leq 2\delta+k-\nu$ because $k\nmid\delta$ implies $k\geq 2$.\\
If the first $t$ components of $u_l$ are zero, what changes in the previous estimation for the weight of $v(z)$ is that we have to subtract $n-k+1$ as $G_\mu u_l=0$ but in turn we could add $k-t=k\mu-\delta$ to each of the weights of $v_i$ for $i=l+1,\hdots, l+\mu-1$. Finally, we obtain
\begin{align*}
&wt(v(z))\\
&\geq (n-k)\mu+n(\mu-1)+2\mu-1-(\mu^2-1)k+\delta+1-\delta-1+(k\mu-\delta)(\mu-1).
\end{align*}
Therefore, we need
$n\geq(\mu+1)k-2+\delta-k\mu+\frac{\delta}{\mu+1}=k+\delta-2+\frac{\delta}{\mu+1}$, which is true since $k+\delta-2+\frac{\delta}{\mu+1}\leq k+\frac{4}{3}\delta-2\leq 2\delta+k-\nu$ because $k\nmid\delta$ implies $k\geq 2$.
\end{proof}

\section{Constructions of MDS convolutional codes}\label{sec:constructions}

In this section, we will use the results of the preceding section to obtain two different constructions for MDS convolutional codes.

\subsection{Constructions for $\delta<k$}

\begin{theorem}[Construction 1] \label{th3}
Assume $\delta<k$, $t$ and $\nu$ as defined before and
$n\geq k+\delta-1$. Moreover, let $\mathbb{F}$ be a finite field with $|\mathbb F|=q$ where $q$ is an odd number such that $q \geq 2 \max\{k+\delta,n\} + 1$ and let $C = \big[ \, c_{ij} \, \big]$ be the Cauchy circulant matrix defined in Theorem \ref{cauchy}.
Set
\begin{equation} \label{eq30}
 \mathcal{G}=\left[\begin{array}{ccc}c_{0,0}& \cdots & c_{0,k+\delta-1}\\\vdots& & \vdots\\ c_{n-1,0} & \cdots & c_{n-1,k+\delta-1} \end{array}\right]
\end{equation}
Then, the matrix $\mathcal{G}$ is superregular. Let us write
$$\mathcal{G}= \left[
          \begin{array}{cccccc}
            g_{0,1} \cdots  g_{\nu,1} & \cdots & g_{0,t}  \cdots g_{\nu,t} & g_{0,t+1}  \cdots g_{\nu-1,t+1} &  \cdots & g_{0,k}  \cdots  g_{\nu-1,k}
          \end{array}
        \right].$$
    Then, $G(z)=\sum_{i=0}^{\mu}G_i z^{i}$ with $G_i=[g_{i,1} \cdots g_{i,k}]$ is the generator matrix of an $(n,k,\delta)$ MDS convolutional code.
\end{theorem}

\begin{proof}
The preceding theorem is an immediate consequence of Theorem \ref{cauchy}.
\end{proof}

\begin{theorem}[Construction 2]
Assume that $\delta < k$, $n\geq k+\delta-1$ and $N\geq 2^{n+k+\delta-1}$.
Let $\alpha$ be a primitive element of a finite field $\mathbb F_{p^N}$.
Set
$$\mathcal{G}=\left[\begin{array}{ccc} \alpha & \cdots & \alpha^{2^{k+\delta-1}}\\ \vdots & & \vdots\\ \alpha^{2^{n-1}} &\cdots & \alpha^{2^{n+k+\delta-2}}\end{array}\right].$$
Then, the matrix $\mathcal{G}$ is superregular. Let us write
$$\mathcal{G}= \left[
          \begin{array}{cccccc}
            g_{0,1} \cdots  g_{\nu,1} & \cdots & g_{0,t}  \cdots g_{\nu,t} & g_{0,t+1}  \cdots g_{\nu-1,t+1} &  \cdots & g_{0,k}  \cdots  g_{\nu-1,k}
          \end{array}
        \right].$$
    Then, $G(z)=\sum_{i=0}^{\mu}G_i z^{i}$ with $G_i=[g_{i,1} \cdots g_{i,k}]$ is the generator matrix of an $(n,k,\delta)$ MDS convolutional code over $\mathbb F_{p^N}$.
\end{theorem}

\begin{proof}
According to Theorem \ref{diagonal}, $\mathcal{G}$ is superregular over $\mathbb F_{p^N}$ if $N$ is greater than $\sum_{i=0}^{k+\delta-1}2^{k+\delta+n-2-2i}=
2^{n-k-\delta}\sum_{i=0}^{k+\delta-1}2^{2i}<
2^{n+\delta+k-1}$. For the last inequality, we used the geometric sum.
\end{proof}

\subsection{Constructions for $\delta\geq k$}

\begin{theorem}[Construction 1] \label{th3}
Assume $\delta\geq k$, $t$ and $\nu$ as defined before and
$n\geq k+2\delta-\nu$.
Moreover, let $\mathbb{F}$ be a finite field with $|\mathbb F|=q$ where $q$ is an odd number such that $q \ge 2n (\nu+1) + 1$ and let $C = \big[ \, c_{ij} \, \big]$ be the Cauchy circulant matrix defined in Theorem \ref{cauchy}.
Set
\begin{equation} \label{eq30}
  g_{j,r}
  =
  \left[
    \begin{array}{c}
      c_{jn,r-1}   \\
      c_{jn+1,r-1} \\
      \vdots       \\
      c_{(j+1)n-1,r-1}
    \end{array}
  \right]
\end{equation}
for $(j,r) \in (\{0,1,\ldots,\nu-1\} \times \{1,2,\ldots,k\}) \cup (\{\nu\} \times \{1,2,\ldots,t\})$ if $t\neq 0$ and for $(j,r) \in (\{0,1,\ldots,\nu-1\} \times \{1,2,\ldots,k\})$ if $t=0$.
Then, the matrix $G(z)=\sum_{i=0}^{\mu}G_i z^{i}$ with $G_i=[g_{i,1} \cdots g_{i,k}]$ is the generator matrix of an $(n,k,\delta)$ MDS convolutional code.
\end{theorem}

\begin{proof}
By Theorem \ref{cauchy}, $C$ is a superregular matrix. Then the matrix $\bar{\mathcal{G}}$ is superregular because it is a submatrix of $C$.
Since $\alpha^{\frac{q-1}{2}} = 1$, we obtain
\[
  c_{u,v}
  =
  \frac{1}{1 - b \alpha^{v-u}}
  =
  \frac{1}{1 - b \alpha^{\frac{q-1}{2}-u+v}}
  =
  c_{0,\frac{q-1}{2}-u+v},
\]
for $0 \leq u,v \leq \frac{q-3}{2}$, and, hence,
\[
  g_{j,r}
  =  \left[
        \begin{array}{c}
          c_{0,\frac{q-1}{2} - jn + r-1}   \\
          c_{1,\frac{q-1}{2} - jn + r-1}   \\
          \vdots                           \\
          c_{n-1,\frac{q-1}{2} - jn + r-1}
        \end{array}
      \right].
\]
Consequently, after an appropriate rearrangement of the columns of $\mathcal{G}$, we obtain a submatrix of the Cauchy matrix $C$.
Therefore, the matrix $\mathcal{G}$ is also superregular.
\end{proof}

\begin{theorem}[Construction 2]
Assume that $\delta\geq k$ and $n\geq k+2\delta-\nu$
and $N\geq 2^{\left(\lfloor\frac{\delta}{k}\rfloor+1\right)n+k-1}$.
Let $\alpha$ be a primitive element of a finite field $\mathbb F_{p^N}$.\\
Set $g_{j,r}=\begin{pmatrix}\alpha^{2^{r-1+nj}}\\ \vdots\\ \alpha^{2^{r-1+n(j+1)}}\end{pmatrix}$ for $r=1,\hdots, k$ and $j=0,\hdots,\lfloor\frac{\delta}{k}\rfloor$ and\\  $g_{\lfloor\frac{\delta}{k}\rfloor+1,r}=\begin{pmatrix}\alpha^{2^{nr-1}}\\ \vdots\\ \alpha^{2^{n(r+1)-2}}\end{pmatrix}$ for $r=1,\hdots,t$.
Then, $G(z)=\sum_{i=0}^{\mu}G_i z^{i}$ with $G_i=[g_{i,1} \cdots g_{i,k}]$ is the generator matrix of an $(n,k,\delta)$ MDS convolutional code over $\mathbb F_{p^N}$.
\end{theorem}

\begin{proof}
With the definitions of the above theorem, $\mathcal{G}$ consists of $k+\delta$ columns of $\left(                                                                            \begin{array}{ccc}
\alpha & \cdots & \alpha^{2^{k-1+n\lfloor\frac{\delta}{k}\rfloor}}\\
\vdots & & \vdots\\
\alpha^{2^{n-1}} & \cdots & \alpha^{2^{k+n-2+n\lfloor\frac{\delta}{k}\rfloor}}
 \end{array}                                                                       \right)$.
Hence, according to Theorem \ref{diagonal}, it is superregular over $\mathbb F_{p^N}$ if $N$ is greater than $\sum_{i=0}^{k+\delta-1}2^{k+n-2+n\lfloor\frac{\delta}{k}\rfloor-2i}=
2^{n+n\lfloor\frac{\delta}{k}\rfloor-k-2\delta}\sum_{i=0}^{k+\delta-1}2^{2i}<
2^{n+n\lfloor\frac{\delta}{k}\rfloor+k-1}$. For the last inequality, we used the geometric sum.
Moreover, $\bar{\mathcal{G}}$ is equal to
$\left(\begin{array}{ccc}
\alpha & \cdots & \alpha^{2^{k-1}}\\
\vdots & & \vdots\\
\alpha^{2^{n-1+n\lfloor\frac{\delta}{k}\rfloor}} & \cdots & \alpha^{2^{k+n-2+n\lfloor\frac{\delta}{k}\rfloor}}
 \end{array}\right)$,
 which, according to Theorem \ref{diagonal}, is superregular over $\mathbb F_{p^N}$ again if $N$ is greater than $\sum_{i=0}^{k+\delta-1}2^{k+n-2+n\lfloor\frac{\delta}{k}\rfloor-2i}<
2^{n+n\lfloor\frac{\delta}{k}\rfloor+k-1}$.                                          \end{proof}

\section{Comparison of constructions for MDS convolutional codes}\label{sec:comparison}

In this section, we want to compare the new constructions for MDS convolutional codes in this paper with the already known constructions. The comparison should be in terms of conditions on the parameters $n,k$ and $\delta$ and in terms of the necessary field size. Throughout this section, we refer to the new constructions of the preceding section as Construction 1 and Construction 2.\\

The constructions in \cite{ju75}, \cite{sm98p1} and \cite{gll06}, which we already mentioned in the introduction, work only for $k=1$ but in this case the required field sizes are smaller than the field sizes required for Construction 1 and Construction 2.\\

For nearly all parameters with $k>1$, the construction of \cite{sm01a} leads to the smallest field size of all known constructions. But this construction has the drawback that it only works for $|\mathbb F|\equiv 1 \mod n$.

Moreover, Construction 1 obtained in this paper could improve the necessary field size of \cite{sm01a} in some particular cases, e.g. it leads to smaller field sizes for $(17,2,1)$ and $(17,2,4)$ convolutional codes.  However, also this construction has restrictions, i.e. it works only for odd field sizes and if $n$ is larger than a particular lower bound.\\

Maximum distance profile (MDP) convolutional codes are convolutional codes whose so-called column distances increase as rapidly as possible for as long as possible; see \cite{mdp} or \cite{strongly} for more explanation.
 As each $(n,k,\delta)$ MDP convolutional code with $(n-k)\mid\delta$ is an MDS convolutional code \cite{strongly}, for comparison, one also has to consider constructions for MDP convolutional codes if $(n-k)\mid\delta$. In \cite{dr13} and \cite[Theorem 3.2]{cmdp}, one could find such constructions that have no other requirements on the parameters than $(n-k)\mid\delta$. There, the required field sizes are larger than the field size from \cite{sm01a} but again this construction has the drawback that it only works for $|\mathbb F|\equiv 1 \mod n$.

 Theorem 3.2 of \cite{cmdp} provides a construction for MDP convolutional codes where the required field size is smaller than the field size in \cite{dr13}. However, it only works for very large characteristic of the field, while the construction in \cite{dr13} and also Construction 2 work for arbitrary characteristic.

If $n$ is sufficiently large, such that the conditions for Construction 2 are fulfilled, it depends on the parameters if it is better than the construction in \cite{dr13} or not. For example, for an $(5,2,2)$ code the construction from \cite{dr13} is the better, and for an $(5,1,5)$ code, Construction 2 is better.

%
%
%
%

\bibliography{mybibfile}

\end{document}